\def\d{\delta}\def\e{\epsilon}
\def\f{\phi}\def\g{\gamma}
\def\k{\kappa}\def\l{\lambda}\def\m{\mu}\def\n{\nu}\def
\p{\pi}\def\q{\psi}\def\r{\rho}
\def\y{\eta}

\def\de{\partial}
\def\inf{\infty}\def\id{\equiv}\def\mo{{-1}}\def\ha{{1\over 2}}

\def\({\left(}\def\){\right)}\def\[{\left[}\def\]{\right]}

\def\Ei{{\rm Ei}}\def\ex{\mathop{\rm e}\nolimits}

\def\erf{\mathop{\rm erf}\nolimits}
\def\erfc{\mathop{\rm erfc}\nolimits}

\def\mn{{\mu\nu}}

\def\tran{transformations }\def\coo{coordinates }

\def\rep{representation }

\def\ms{maximally symmetric }

\def\poi{Poincar\'e }

\def\wrt{with respect to }\def\ie{i.e.\ }

\def\kp{$\k$-\poi }

\def\section#1{\bigskip\noindent{\bf#1}\smallskip}

\def\PL#1{Phys.\ Lett.\ {\bf#1}}
\def\PRL#1{Phys.\ Rev.\ Lett.\ {\bf#1}}
\def\PR#1{Phys.\ Rev.\ {\bf#1}}
\def\NP#1{Nucl.\ Phys.\ {\bf#1}}
\def\JMP#1{J.\ Math.\ Phys.\ {\bf#1}}

 \def\IJMP#1{Int.\ J. Mod.\ Phys.\ {\bf #1}}
 
\def\PRep#1{Phys.\ Rep.\ {\bf#1}}

\def\EPJ#1{Eur.\ Phys.\ J.\ {\bf#1}}

\def\hep#1{{\tt hep-th/#1}}\def\arx#1{{\tt arXiv:#1}}
\def\PRep#1{Phys.\ Rep.\ {\bf#1}}

\def\ref#1{\medskip\everypar={\hangindent 2\parindent}#1}
\def\beginref{\begingroup
\bigskip
\centerline{\bf References}
\nobreak\noindent}
\def\endref{\par\endgroup}

\def\erf{\mathop{\rm erf}\nolimits}\def\cD{{\cal D}}
\def\ms{\(1-{p_0\over\k}\)}

\magnification=1200

{\nopagenumbers
\line{}
\vskip30pt
\centerline{\bf Vacuum energy from noncommutative models}

\vskip60pt
\centerline{
{\bf S. Mignemi}$^{1,2,}$\footnote{$^\ddagger$}{e-mail: smignemi@unica.it},
and {\bf A. Samsarov}$^{3,}$\footnote{$^*$}{e-mail: andjelo.samsarov@irb.hr},}
\vskip10pt
\centerline{$^1$Dipartimento di Matematica e Informatica, Universit\`a di Cagliari}
\centerline{viale Merello 92, 09123 Cagliari, Italy}
\smallskip
\centerline{$^2$INFN, Sezione di Cagliari, Cittadella Universitaria, 09042 Monserrato, Italy}
\smallskip
\centerline{$^3$Rudjer Bo\v skovi\'c Institute, Bijeni\v cka cesta 54, 10002 Zagreb, Croatia}
\vskip80pt

\centerline{\bf Abstract}
\medskip
{\noindent The vacuum energy is computed for a scalar field in a noncommutative background in several models
of noncommutative geometry. One may expect that the noncommutativity introduces a natural cutoff on the ultraviolet
divergences of field theory. Our calculations show however that this depends on the particular model considered:
in some cases the divergences are suppressed and the vacuum energy is only logarithmically divergent, in other
cases they are stronger than in the commutative theory.
}
\vskip10pt
{\noindent

}
\vskip80pt\
\vfil\eject}

\section{1. Introduction}
Noncommutative models [1,2] may improve the behavior of field theories in the ultraviolet region by smoothing or
removing some of the singularities of commutative quantum field theory. In fact, they imply a lower bound on the
length scales, given by the inverse of the noncommutativity parameter $\k$, or equivalently an upper bound on the
energy scales, given by $\k$. The scale $\k$ is usually assumed to of the order of the Planck mass $M_P\sim 10^{19}$ GeV
(but lower values are not excluded), and may act as a natural cutoff on the
divergences of quantum field theory, in contrast with the commutative case where the cutoff must be imposed by hand.

This possibility may be tested in the calculation of the vacuum energy of quantum fields. This computation has interesting
implications on cosmology, since the vacuum energy is often identified with the cosmological constant [3].
Although this argument is almost certainly wrong, since it is not based on a well-defined theory and predicts a value that
can be 120 orders of magnitudes greater than the observed one, it can still be interesting to check if the noncommutativity
parameter can act as a natural cutoff and improve the ultraviolet behavior of the theory. Of course, if $\k$ is of Planck
scale, this does not change much the predictions from a phenomenological point of view, since in this context also the
standard UV cutoff is usually assumed to have the same scale.

In this paper, we calculate the vacuum energy of a massless scalar field in noncommutative background, using the heat kernel
method. This method allows to evaluate the one-loop effective action by calculating the integral of an operator, related
to the solution of the heat equation on a Euclidean manifold.
We shall follow the approach of [4], where models presenting a breaking of Lorentz invariance are studied.
A review of the heat kernel formalism can be found for example in [5].
A calculation similar to the present one, but differing in several respects and based on a perturbative expansion in the
noncommutativity parameter, has also been performed in [6].

We investigate a class of noncommutative models characterized by a deformation of the Heisenberg algebra, which in turn
implies a deformation of the \poi symmetry and hence of the field equations.
Also the measure of the Hilbert space must be adapted to the nontrivial \rep of the deformed Heisenberg algebra, and these
two effects combine to modify the value of the heat kernel integral in comparison with the commutative one.

We show that,
contrary to naive expectations, noncommutativity does not completely regularize the theory, and only in some of the models
examined the UV behavior is improved \wrt the commutative theory, while in other models it can be worsened.
The best improvement occurs in the anti-Snyder model, where the trace of the heat kernel is finite, and the divergence
of the vacuum energy is only logarithmic.

\section{2. Heat kernel}
Let us consider a field theory obeying the equation\footnote{$^1$}{We adopt the following conventions: metric  $\y_\mn=
{\rm diag}(-1,1,1,1)$; $\m=0,1,2,3$; $i=1,2,3$; $v^2=v^\m v_\m$.}
$$\cD\f=F(\de_0,\de_i)\f=0,\eqno(1)$$
where $\cD$ is a differential operator that deforms the usual Laplacian $\de^\m\de_\m$ by a
parameter $\k$, in such a way to preserve the invariance under spatial rotations.

It is known that for a quantum bosonic field in Euclidean space, with partition function defined as $Z=(\det\cD)^{-1/2}$,
the one-loop effective action $W=\ha\ln\det\cD$ can be written in terms of the heat kernel,
$$W=-\ha\int_{1/\e^2}^\inf\ {ds\over s}\,K(s),\eqno(2)$$
where $s$ is a real parameter and $K(s)=\int dx<x\,|\ex^{-s\cD}|\,x>$ is the trace of the
heat kernel.
The cutoff $1/\e^2$, with $\e\gg1$, at the lower limit is introduced because in standard field theory the integral (2) is
usually divergent for $s\to0$ (UV divergence).
We recall that the heat kernel $K(s,x,x')=\ <x\,|\ex^{-s\cD}|\,x'>$ is defined as a solution of the heat equation
$$(\de_s+\cD)K(s,x,x')=0,\qquad K(0,x,x')=\d(x,x').\eqno(3)$$

The calculations are most easily performed in momentum space, where the solution of the heat equation is
trivial. It follows that [4]
$$K(s)={V\over(2\p)^d}\int_{-\inf}^\inf d^dp\ \ex^{-sF(p_0,p_i)},\eqno(4)$$
where $V$ is the volume of spacetime.
In the special case of the undeformed Laplace operator, $K(s)=V/(4\p s)^{d/2}$.

The effective action follows from eq.\ (2). The vacuum energy density $\l$ is defined as
$$\l=-{W\over V},\eqno(5)$$
and $\l$ is often identified with the cosmological constant. Of course, in standard field theory the value of $\l$ depends
on the cutoff $\e$ introduced to regularize the UV divergences of the effective action. In particular, in four dimensions
$\l=\e^4/64\p^2$. One may hope that in noncommutative
models these divergences might be regularized by the noncommutativity scale $\k$, so that the calculation gives a finite
result without need of introducing an artificial cutoff.
We want to study if this happens in some well-known cases. For ease of calculation, we consider massless fields, that may
however lead to IR divergences. We shall always understand that these are regularized when one considers massive fields.

\section{3. Noncommutative models}
Noncommutative theories are based on the hypothesis that spacetime has a granular structure, implemented through the noncommutativity
of spacetime coordinates, with a scale of length $\k^\mo$, that is usually (but not necessarily) identified with the Planck length.
Because of the presence of this fundamental scale, most noncommutative geometries are associated to a deformation of the action of
Lorentz \tran on phase space, and hence of the \poi algebra.

It must be noted that their properties are not completely determined by the noncommutativity of spacetime coordinates,
since the same noncommutative \coo can be associated with different coproducts.
Different realizations of a given noncommutative geometry are often called bases and usually lead to different physical predictions.
The different bases can be characterized by specifying their deformed Heisenberg algebra, generated by the position operators $x_\m$
and the momentum operators $p_\m$. From the knowledge of this algebra, one can obtain the coproduct of momenta and the other
relevant quantities.

In order to calculate the heat kernel one must first of all establish the field equations.
The deformed invariance of the theory is preserved if the (momentum space) deformed Laplace equation is identified with the Casimir
operator $C$ of the deformed \poi algebra. However, this choice is not unique, because any function of $C$ could be adopted.

Moreover, a nontrivial measure must be fixed on the Hilbert space, again invariant under deformed Lorentz transformations.
To single out this measure uniquely, we also require that the position operators are symmetric in the \rep chosen.

In this paper, we consider some specific models that lead to simple calculations of the heat kernel:
the first one is the Snyder model [7]. Its main peculiarity is that it preserves the standard action of the Lorentz group
on phase space, and it can be seen as dual to de Sitter spacetime.
Its deformed Heisenberg algebra, in the original Snyder basis, is given by
$$[x_\m,x_\n]=i{J_\mn\over\k^2},\quad[p_\m,p_\n]=0,\quad[x_\m,p_\n]=i\(\y_\mn+{p_\m p_\n\over\k^2}\),\eqno(6)$$
where $J_\mn=x_\m p_\n-x_\n p_\m$ are the generators of the Lorentz algebra.
Since the Lorentz \tran are not deformed in this case, the Casimir operator is simply given by{\footnote{$^2$}{As noted
above, this choice is not unique. Sometimes the choice $C=p^2/(1-p^2/\k^2)$ is made.}
$$C=p^2\id-p_0^2+p_i^2,\eqno(7)$$
with $m^2=-p^2<\k^2$. This implies an upper bound for the allowed particle masses in this model.

There is also the possibility of choosing the opposite sign in front of $\k^2$ in (6) (anti-Snyder geometry) [7,8]. In this case
there is no upper bound on the particle masses. However, all the relations we shall discuss hold true, by simply changing
the sign in front of $\k^2$.

The other examples belong to the \kp class: one is the so-called Magueijo-Smolin model [9]. The nontrivial commutators of its
Heisenberg algebra in the Granik basis [10] are
$$[x_i,x_0]=i{x_i\over\k},\quad[x_0,p_i]=i{p_i\over\k},\quad[x_i,p_j]=i\d_{ij},\quad[x_0,p_0]=-i\(1-{p_0\over\k}\).\eqno(8)$$
The \poi algebra is now deformed and its Casimir operator is
$$C={-p_0^2+p_i^2\over\ms^2}.\eqno(9)$$
In this case, the bound $p_0<\k$ must hold.

The last one is the Majid-Ruegg (MR) model [11], sometimes called bicrossproduct basis of the \kp model [2]. Its Heisenberg
algebra reads
$$[x_i,x_0]=i{x_i\over\k},\quad[x_0,p_i]=i{p_i\over\k},\quad[x_i,p_j]=i\d_{ij},\quad[x_0,p_0]=-i.\eqno(10)$$
Also in this case the \poi algebra is deformed, with Casimir operator
$$C=-\(2\k\sinh{p_0\over2\k}\)^2+\ex^{p_0\over\k}p_i^2.\eqno(11)$$

The Euclidean version of this class of noncommutative models is usually obtained by an analytic continuation of the
Lorentzian theory, with $p_0\to ip_0$,  $\k\to i\k$, see discussions in [12] for the $\k$-\poi model. This assumption
appears natural considering that $\k$ plays the role of an energy scale. Requiring $\k\to i\k$ also appears necessary
in order to get a physically sensible interpretation of the Euclidean theory in the $\k$-\poi case, while in the Snyder
case the situation is less clear.

For the class of models considered here, the calculation of the heat kernel differs in two ways from the standard
case: first, as discussed above, the Laplace operator is chosen proportional to the Casimir operator, in order
to be invariant under the deformed Lorentz transformations. Moreover, the measure of momentum space is not trivial, since
the phase space operators satisfy a deformed Heisenberg algebra, and hence a nontrivial realization on the Hilbert space
must be found.


\section{4. Snyder model}
We define the Euclidean Snyder model by the analytic continuation $p_0\to i p_0$, $\k\to i\k$, as for the \kp models.
Keeping instead $\k\to\k$ would simply interchange the roles of Snyder and anti-Snyder Euclidean spaces. Our choice maintains
the bound $m^2<\k^2$ also in the Euclidean case.

In both instances, the Euclidean Casimir operator is given by
$$C=p_E^2=p_0^2+p_i^2.\eqno(12)$$

We start by considering anti-Snyder space, because it gives rises to more interesting results.
Euclidean anti-Snyder space in the basis (6) can be realized by a suitable choice of operators in a quantum representation.
Two main choices can be found in the literature, in terms of a standard Hilbert space of functions of a canonical momentum
variable $P_\m$, with Euclidean signature:
the first one reads [7]
$$p_\m=P_\m,\qquad  x_\m=i{\de\over\de P_\m}+{i\over\k^2\,}P_\m P_\n{\de\over\de P_\n},\eqno(13)$$
with $-\inf<P_\m<\inf$.
We require that in this representation the position operators $ x_\m$ be symmetric, \ie that $<\q\;| x_\m|\;\f>$ =
$<\f\;| x_\m|\;\q>$.
This occurs if one introduces  a nontrivial measure in the $P$-space [13],
$$d\m={d^4P\over(1+P^2/\k^2)^{d+1\over2}},\eqno(14)$$
with $d$ the dimension of the space.

We can now proceed to compute the trace of the heat kernel. In this representation, the Laplacian (12) is simply given by $P^2$.
Hence, for $d=4$, eq.\ (4) gives
$$K(s)={V\over 16\p^4}\int_{-\inf}^\inf{d^4P\over(1+P^2/\k^2)^{5/2}}\ \ex^{-sP^2}.\eqno(15)$$
Defining polar coordinates, with radial coordinate $\r=\sqrt{P^2}$, after integrating on the angular variables, (15) becomes
$$K(s)={V\over8\p^2}\int_0^\inf{\r^3d\r\over(1+\r^2/\k^2)^{5/2}}\ \ex^{-s\r^2}.\eqno(16)$$
Performing the integral, one obtains
$$K(s)={\k^4 V\over24\p^2}\[2(1+ \k^2 s)-\k\sqrt{\p s}\,(3 + 2\k^2s)\,\ex^{\k^2 s}\erfc(\k\sqrt s)\],\eqno(17)$$
where $\erfc(x)$ is the complementary error function, $\erfc(x)={2\over\sqrt\p}\int_x^\inf\ex^{-t^2}dt$.

In the limit $s\to0$, $K(s)$ takes the finite value ${\k^4V\over12\p^2}$, in contrast with the commutative theory, where it diverges
as $s^{-2}$. Also in the limit  $s\to\inf$ it takes a finite value.
The evaluation of the vacuum energy (5) gives then for $s\to0$ a logarithmic divergence, and the vacuum energy density reads
$$\l={\k^4\over12\p^2}\ \ln{\e\over m},\eqno(18)$$
where $m$ is the IR scale. The divergence is much milder than the commutative result, $\l\sim\e^4$, although, if one assumes
$\e\sim\k$, the numerical value of $\l$ is not much different in the two cases.

\medskip
The same results can be obtained using another \rep of the relations (6), again defined in terms of a standard Hilbert space
of functions of a canonical momentum variable $P_\m$  [8],
$$ p_\m={P_\m\over\sqrt{1-P^2/\k^2}},\qquad x_\m=i\sqrt{1-P^2/\k^2}\,{\de\over\de P_\m},\eqno(19)$$
with $P^2<\k^2$.
In this case, the measure for which the operators $x_\m$ are symmetric is given by [8]
$$d\m={d^4P\over\sqrt{1-P^2/\k^2}},\eqno(20)$$
independently from the dimension of the space, while $p^2={P^2\over 1-P^2/\k^2}$.
Hence,
$$K(s)={V\over16\p^4}\int_{P^2<\k^2}{d^4P\over\sqrt{1-P^2/\k^2}}\ \ex^{-{sP^2\over1-P^2/\k^2}}.\eqno(21)$$
In polar \coo $\r=\sqrt{P^2}$, this becomes after integration on the angular \coo
$$K(s)={V\over8\p^2}\int_0^{\k^2}{\r^3 d\r\over\sqrt{1-\r^2/\k^2}}\ \ex^{-{s\r^2\over1-\r^2/\k^2}}.\eqno(22)$$
By a change of variables $\r\to {\r\over\sqrt{1-\r^2/\k^2}}$, one finally recovers (16).
\medskip

The Snyder model is obtained by replacing $\k^2$ with $-\k^2$ in (13) and (14), but now the
calculation is more involved, since the integral for $K(s)$ does not converge on the boundary.
In fact, the \rep (13) becomes now
$$p_\m=P_\m,\qquad  x_\m=i{\de\over\de P_\m}-{i\over\k^2\,}P_\m P_\n{\de\over\de P_\n},\eqno(23)$$
with $P^2<\k^2$, and measure
$$d\m={d^4P\over(1-P^2/\k^2)^{d+1\over2}}.\eqno(24)$$
The integral (16) becomes
$$K(s)={V\over8\p^2}\int_0^\k{\r^3d\r\over(1-\r^2/\k^2)^{5/2}}\ \ex^{-s\r^2},\eqno(25)$$
that diverges at $\r=\k$. One must therefore introduce an UV cutoff already at this stage, for example taking as
upper limit of integration $\e<\k$. This gives at leading order the constant value
$K(s)\sim{\k^5(\e^2-2\k^2/3)\over(\k^2-\e^2)^{3/2}}$.
Taking the same cutoff $\e$ for the integration over $s$, it follows that
$$\l\sim-{\k^5(\e^2-2\k^2/3)\over(\k^2-\e^2)^{3/2}}\,\ln{\e\over m}.\eqno(26)$$
Therefore, in this case the vacuum energy diverges as $(\k-\e)^{-3/2}$, with $(\k-\e)\ll1$.

\section{5. MS model}
This model belongs to the \kp class and considerations analogous to those of [12] suggest that the Euclidean theory
can be defined through the prescription $p_0\to ip_0$, $\k\to i\k$.
This leads to the Euclidean Laplacian
$$C={p_0^2+p_i^2\over\ms^2}.\eqno(27)$$
The action of the 4-dimensional rotations on phase space is deformed and only the action of the spatial rotations
is preserved.
However, the calculation can be performed in a way similar to the one of the previous section.

First of all, we notice that the MS model can be represented on a standard Hilbert space of functions of
a canonical momentum variable $P_\m$ as
$$ p_\m={P_\m\over1+P_0/\k},\qquad x_\m=i(1+P_0/\k)\,{\de\over\de P_\m},\eqno(28)$$
where $-\inf<P_i<\inf$, $0<P_0<\inf$.
In this representation, the measure for which the operators $ x_\m$ are symmetric is given by
$$d\m={d^4P\over1+P_0/\k},\eqno(29)$$
The heat kernel integral becomes therefore
$$K={V\over16\p^4}\int {d^4P\over1+P_0/\k}\ \ex^{-sP^2},\eqno(30)$$
and can be separated into
$$K={V\over16\p^4}\int_0^\inf{dP_0\over1+P_0/\k}\ \ex^{-sP_0^2}\int_{-\inf}^\inf d^3P_i\ \ex^{-sP_i^2}.\eqno(31)$$
This gives
$$K=-{\k V\over32\p^{3/2}}\,{\ex^{-\k^2s}\over s^{3/2}}\,[i\p\erf(i\k\sqrt s)+\Ei(\k^2s)],\eqno(32)$$
where erf$(x)$ is the error function and Ei$(x)$ the exponential integral.

For $s\to0$, $K\sim-{\k V\over32\p^{3/2}}s^{-3/2}\big(\ln(\k^2s)+\g+O(s)\big)$, where $\g$ is the Euler-Mascheroni constant.
For $s\to\inf$, $K$ vanishes.
As one could have guessed from the structure of the integral, in this case only the integration on $p_0$
gives rise to a milder UV divergence, while the spatial part presents the usual divergence $s^{-3/2}$, with a
further logarithmic factor.
The calculation of the vacuum energy density gives at leading order
$$\l\sim{1\over24\p^{3/2}}\,\k\e^3\ln{\e\over\k}.\eqno(33)$$
The UV divergence is milder than in the commutative case. With the natural identification $\e=\k$, the logarithmic term
vanishes, and the leading divergence is given by the next term in the expansion, with the standard $\k^4$ behavior.

\section{6. MR model}
Let us consider now the MR model. As discussed before, the Euclidean theory is obtained
for $p_0\to ip_0$, $\k\to i\k$.
The Euclidean Laplacian is then
$$C=\(2\k\sinh{p_0\over2\k}\)^2+\ex^{p_0\over\k}p_i^2.\eqno(34)$$
A representation of the Heisenberg algebra is given by
$$ p_\m=P_\m,\qquad  x_0=i{\de\over\de P_0}-{i\over\k\,}P_i{\de\over\de P_i},\qquad x_i=i{\de\over\de P_i}.\eqno(35)$$
These operators are Hermitian for the measure [14]
$$d\m=\ex^{3P_0\over\k}\,d^4P,\eqno(36)$$
The heat kernel integral becomes then
$$K(s)={V\over16\p^4}\int_{-\inf}^\inf d^4P\,\ex^{3P_0\over\k}\ex^{-s\[4\k^2\sinh^2{P_0\over2\k}+\,\ex^{P_0/\k}P_i^2\]},\eqno(37)$$
or
$$K(s)={V\over16\p^4}\int_{-\inf}^\inf dP_0\,\ex^{3P_0\over\k}\ex^{-4\k^2s\,\sinh^2{P_0\over2\k}}
\int_{-\inf}^\inf d^3P_i\,\ex^{-s\ex^{P_0/\k}P_i^2},\eqno(38)$$
and, after integration over the spatial coordinates,
$$K(s)={V\over16\p^{3/2}}\,{1\over s^{3/2}}\int_{-\inf}^\inf dP_0\,\ex^{3P_0\over2\k}\,\ex^{-4\k^2s\,\sinh^2{P_0\over2\k}}.\eqno(39)$$
The last integration gives
$$K(s)={V\over32\p^2\k^2s^3}\,(1+2\k^2s).\eqno(40)$$
In this case, the divergence for $s\to0$ is worse than in the commutative case, while the expression converges for $s\to\inf$.

Computing the vacuum energy density we obtain
$$\l={\e^6\over192\p^2\k^2}\(1+{3\k^2\over\e^2}\).\eqno(41)$$
Again, if one identifies $\e$ with $\k$, $\l\propto\k^4$, like in the standard theory.
\section{7. Conclusions}
Using the heat kernel method, we have shown that in some cases noncommutativity can regularize the behavior of the vacuum energy of a
scalar field theory.
This is however not a universal property: it holds for the anti-Snyder model, but not necessarily in different instances, like the MR
or the MS model.
It is important to remark that our results are independent of the representation chosen in a Hilbert space. This has been shown
explicitly for the Snyder model, but can be checked also in the other cases. It is however crucial to choose the correct measure in the
Hilbert space.

The results obtained here are in agreement with explicit calculations of quantum field theory in Snyder space, which show an improvement
of the divergences with respect to the commutative case [15]. Analogous conclusions concerning the energy of the vacuum in
noncommutative theories have been obtained using a very different approach related to the Wheeler-deWitt equation, in ref.\ [16].

\bigskip
\section{Acknowledgements}
A.S. would like to thank the University of Cagliari and INFN, Sezione di Cagliari, for their kind hospitality, and
acknowledges support by the Croatian Science Foundation under the project (IP-2014-09-9582) and a partial support by the H2020
CSA Twinning project No. 692194, RBI-T-WINNING.
\beginref
\ref [1] S. Doplicher, K. Fredenhagen and J.E. Roberts, \PL{B331}, 39 (1994).
\ref [2] J. Lukierski, H. Ruegg, A. Novicki and V.N. Tolstoi, \PL{B264}, 331 (1991);
J. Lukierski, A. Novicki and H. Ruegg, \PL{B293}, 344 (1992).
\ref [3] An historical review is given in S.E. Rugh and H. Zinkernagel, \hep{0012253} (2000).
\ref [4] D. Nesterov and S.N. Solodukhin, \NP{B842}, 141 (2017).
\ref [5] D.V. Vassilevich, \PRep{388}, 279 (2003);
D. Fursaev and D. Vassilevich, {\it Operators, Geometry and Quanta, Methods of Spectral
Geometry in Quantum Field Theory}, Springer, New York, 2011.
\ref [6] A. Samsarov, \arx{1711.10894}.
\ref [7] H.S. Snyder, \PR{71}, 38 (1947).
\ref [8] S. Mignemi, \PR{D84}, 025021 (2011).
\ref [9] J. Magueijo and L. Smolin, \PRL{88}, 190403 (2002).
\ref [10] A. Granik, \hep{0207113}.
\ref [11] S. Majid and H. Ruegg, \PL{B334}, 348 (1994).
\ref [12] J. Lukierski and H. Ruegg, \JMP{35}, 2607 (1994);
 D. Benedetti, \PRL{102}, 111303 (2009);
 M. Arzano and T. Tre\'skniewski, \PR{D89}, 124024 (2014).
\ref [13] Lei Lu and A. Stern, \NP{B854}, 894 (2011).
\ref [14] G. Amelino-Camelia and S. Majid, \IJMP{A15}, 4301 (2000);
G. Amelino-Camelia, V. Astuti and G. Rosati, \EPJ{C73}, 2521 (2013).
\ref [15] S. Meljanac, S. Mignemi, J. Trampeti\'c and J. You, \arx{1711.09639}.
\ref [16] R. Garattini and P. Nicolini, \PR{D83}, 064021 (2011).
\endref
\end